\documentclass[useAMS,usenatbib]{mn2e}
\usepackage{graphicx,amssymb, txfonts,pstricks}

\title[HCN emission from  G75.78+0.34 and G75.77+0.34]{HCN emission from the H{\,\sc ii} 
regions G75.78+0.34 and G75.77+0.34}
\author[Riffel and L\"udke]{Rogemar. A. Riffel\thanks{E-mail:
rogemar@smail.ufsm.br} and Everton L\"udke
\\
Universidade Federal de Santa Maria, Departamento de F\'\i sica, LARIE, Centro de Ci\^encias Naturais e Exatas, 97105-900, Santa Maria, RS, Brazil \\ 
}

\begin{document}

\date{Accepted 1988 December 15. Received 1988 December 14; in original form 1988 October 11}

\pagerange{\pageref{firstpage}--\pageref{lastpage}} \pubyear{2002}

\maketitle

\label{firstpage}

\begin{abstract}
We present images for the 3.5~mm continuum and HCN(J=1--0) hyperfine line emission from the surroundings of the H\,{\sc ii} 
regions G75.78+0.34 and G75.77+0.34 obtained with the Berkeley Illinois Maryland
Association (BIMA) interferometer using the D configuration at a spatial resolution of $\sim$18 arcsec 
and spectral sampling of 0.34 km\,s$^{-1}$. 
The continuum emission of both objects is dominated by free-free emission
from the ionized gas surrounding the exciting stars. Dust emission may contribute only a small fraction of the 3.5\,mm continuum
 from  G75.78+0.34 and is negligible for G75.77+0.34. 
The high spectral resolution reached by BIMA allowed us to separate the emission from each hyperfine transition (F=1-1, F=2-1 and F=0-1),
 as well as to construct velocity channel maps along each emission-line profile. 
The HCN flux distributions are similar to those observed for the CO emission, 
but with some knots of high intensities indicating that the HCN traces high density clouds not seen in CO. 
The HCN hyperfine line ratios for both H\,{\sc ii} regions differ from those predicted theoretically for Local Thermodynamic Equilibrium (LTE),
 probably due to scattering of radiation processes. The velocity channels 
shows that the HCN emission of G75.78+0.34  follows the bipolar molecular outflows previously observed in CO. For G75.77+0.34, the outflowing 
gas contributes only a small fraction of the HCN emission.
\end{abstract}


\section{Introduction}

H\,{\sc ii} regions are among the most well-studied class of objects of our Galaxy. However 
 most of these studies are based on optical imaging and spectroscopy, in which 
the earliest stages of the life of stars cannot be accessed. At the early stages, the star(s) and  
the surrounding H\,{\sc ii} region, are still embedded in the molecular cloud which is 
 invisible at optical wavelengths. Nevertheless,
they can be observed at longer wavelength, such as radio 
and infrared bands, which are less affected by extinction
than the optical region of the spectra \citep{wood89,shepherd97,carral97,franco00a,roman-lopes09}.

    Usually, according to their sizes, density, mass of ionized 
gas and emission measure (EM), H\,{\sc ii} regions are classified as ultracompact, compact 
and extended \citep[e.g.][]{habing79}.
 Ultracompact H\,{\sc ii} regions have sizes of $\lesssim$0.1\,pc, display 
densities of $\gtrsim$10$^4$\,cm$^{−3}$ , mass of ionized gas of $\sim$10$^{-2}$\,M$_\odot$
 and EM$\gtrsim$10$^7$\,pc\,cm$^{-6}$. They are located in the inner, high-pressure,
regions of molecular clouds \citep[e.g.][]{kurtz94,franco00b}.
 Compact H\,{\sc ii} regions have sizes of
  $\lesssim$0.5\,pc, densities $\gtrsim5\times$10$^3$\,cm$^{−3}$, mass of ionized gas of $\sim$1\,M$_\odot$
and EM$\gtrsim$10$^7$\,pc\,cm$^{-6}$, while classical H\,{\sc ii} regions have sizes
$\sim$10\,pc, densities of $\sim$100\,cm$^{−3}$, $\sim$10$^5$\,M$_\odot$ of ionized gas and
EM$\sim$10$^2$\,pc\,cm$^{−6}$ \citep{franco00b,kurtz05}. \citet{franco00b} 
 present an expansion of this classification including 
sub-classes and discuss the physical properties of each class
\citep[see also ][for a review of the physical properties of each class]{kurtz05}.

    The study of the molecular emission close to H\,{\sc ii} regions
is a key to understand the physical properties of
these objects and how stars form. These studies are often based
on observations of CO emission for low rotational microwave transitions
\citep[e.g.][]{matthews86,shepherd96,shepherd97,qin08}.
 The low dipole moment of CO implies that low rotational transitions do not trace dense gas, while
molecules with higher dipole moments can be used to observe
high density gas. A typical tracer of the emission from dense
cores ($n\sim10^4 - 10^5\,{\rm~cm^{−3}}$) is the J=1--0 transition from the HCN
molecule \citep[e.g.][]{afonso98}.

     In this work we present observations of HCN and 3.5~mm
continuum images for the H\,{\sc ii}  regions G75.78+0.34 and
G75.77+0.34 obtained with the Berkeley Illinois Maryland
Association (BIMA) interferometer. These objects were selected because 
they present a well-known molecular outflow observed in  the
CO(J=1--0) emission  \citep{shepherd96,shepherd97}, thus being 
ideal candidates to investigate on whether the high-density gas traced by the HCN 
has the same distribution and kinematics as the low-density gas traced by CO. 
These objects are localized in the giant molecular
cloud ON2 and were firstly identified by \citet{matthews73}
 with observations at 5 and 10.7 GHz. These H\,{\sc ii}  regions
are known to be located at a distance of 5.5 kpc \citep{wood89}.
G75.77+0.34 presents typical parameters of a compact H\,{\sc ii}  
region and it is excited by an O star, while G75.78+0.34 is at an earlier 
stage of evolution, being classified as an ultracompact H\,{\sc ii}  
region and is excited by a B star \citep{matthews73,wood89,shepherd96,carral97,franco00a,kurtz05}.
Previous studies of these regions include also the identification
of several H$_2$O maser sources close to the positions of both H\,{\sc ii}
regions \citep{hofner96} as well as millimetric radio sources \citep{carral97}.

    The main goal of this work is to map the distribution and
kinematics of the HCN emitting gas from the surroundings of both H\,{\sc ii} regions,  
 and compare with those of the CO emitting gas.
 A secondary goal is to
map and discuss the origin of the 3.5~mm radio continuum emission
from both regions.
    In Section 2 we describe the observations and the data reduction.
 The results for the 3.5~mm continuum and HCN line emission are presented in Section 3,
 while the discussion of these
results are presented in Section 4. Section 5 presents the conclusions of this work.

\section{Observations and data reduction} \label{obs}

\begin{figure}
\centering
\includegraphics[scale=0.43]{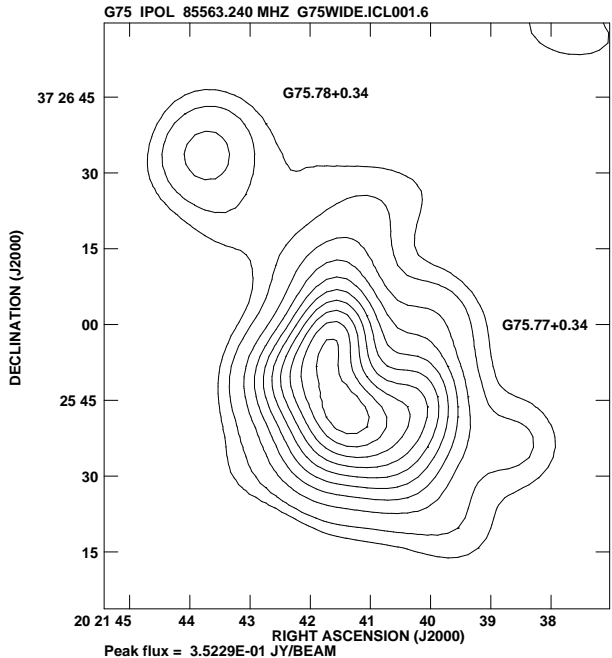}
\caption{ 3.5\,mm emission map of G75.78+0.34 and G75.77+0.34 obtained with BIMA at D configuration. 
The contour levels shown are: $35\times[1, 2, 3, 4, 5, 6, 7, 8, 9, 10]$\,mJy.} 
\label{continuum}
\end{figure}

Interferometric observations of  G75.78+0.34 and G75.77+0.34 
were obtained with the Berkeley Illinois Maryland Association (BIMA) in June, 1999. 
A detailed description of the BIMA interferometer can be found in \citet{welch96}. 
We used the shortest baseline configuration (D-array), with maximum baseline length of 8900\,k$\lambda$,
 in order to observe the HCN(J=1--0) emission  at 88.63\,GHz, as well as the 3.5\,mm continuum emission. 

The primary amplitude and bandpass calibrators were Mars and 3C273, respectively, the latter with a flux density of 22.5 Jy at 3.5~mm. 
3C454.3 was also used as a control amplitude calibrator with a derived flux density of 7.2 Jy. An 830-MHz wide continuum 
channel has been used to produce the 3~mm image, which was strong enough to be used to self-calibrate the visibilities.
 
The data reduction was performed using standard procedures with the {\sc miriad} software and included amplitude, 
phase and bandpass calibrations. The calibrated data has been transported to the {\sc aips} software for self-calibration and imaging. 
The resulting full width at half maximum (FWHM) of the synthesized map  was about 18 arcsec and the resulting frequency 
sampling was $\delta\nu\approx97.656$\,kHz for the line emission, corresponding to a velocity sampling of  
$\delta V\approx 0.34\,{\rm km\,s^{-1}}$.

\section{Results} \label{res}

\subsection{The 3.5\,mm continuum emission}

In Figure\,\ref{continuum} we present the the 3.5~mm continuum image obtained
from the 830-MHz wide channel, which shows that
 G75.78+0.34 presents a compact circular shape unresolved by our observations, thus corresponding 
to a linear diameter $\leq$0.5\,pc. On the other hand, G75.77+0.34
present a more extended emission which can be described as a
``curved" structure with size of about 1$^\prime$, corresponding to a linear
diameter of $\sim$1.6 pc.

G75.78+0.34 presents a 3.5\,mm total flux of $\sim$119\,mJy and a peak flux of 92\,mJy beam$^{-1}$ at the position
 $\alpha=$20h\,21m\,43.6s and $\delta=$37$^\circ$\,26$^\prime$\,36$^{\prime\prime}$, approximately coincident with the peak position 
of higher resolution Very Large Array (VLA) images at 6\,cm and 7\,mm \citep{wood89,carral97}. The continuum emission for G75.77+0.34 peaks at 
$\alpha=$20h\,21m\,41.4s and $\delta=$37$^\circ$\,25$^\prime$\,44$^{\prime\prime}$ with a peak flux density of 352\,mJy and 
a total flux density of $\approx$1.73\,Jy. The morphology of the 3.5\,mm emission is similar to the 6\,cm image 
presented by \citet{matthews73} at similar spatial resolution. 

\subsection{The HCN emission}

\begin{table*}
\centering
\caption{Measured fluxes for each transition and flux ratios for G75.78+0.34 and G75.77+0.34. The fluxes are given 
in Jy and the errors are about 10\%. LTE (optically thin) values of $R_{02}$ and $R_{12}$ are 0.2 and 0.6, respectively.}
\begin{tabular}{c c c c c c }
\hline
\hline
Object       & F=1-1 & F=2-1 & F=0-1 & $R_{02}{\rm(F=0-1/F=2-1)}$ &  $R_{12}{\rm(F=1-1/F=2-1)}$ \\
\hline
G75.78+0.34 & 17.47  & 14.89 & 21.25 & 1.43                       & 1.17 \\
G75.77+0.34 & 4.39   & 18.41 & 11.75 & 0.64                       & 0.24 \\
\hline
\end{tabular}
\label{hcn_flux}
\end{table*}

\begin{figure}
\centering
\includegraphics[scale=0.43]{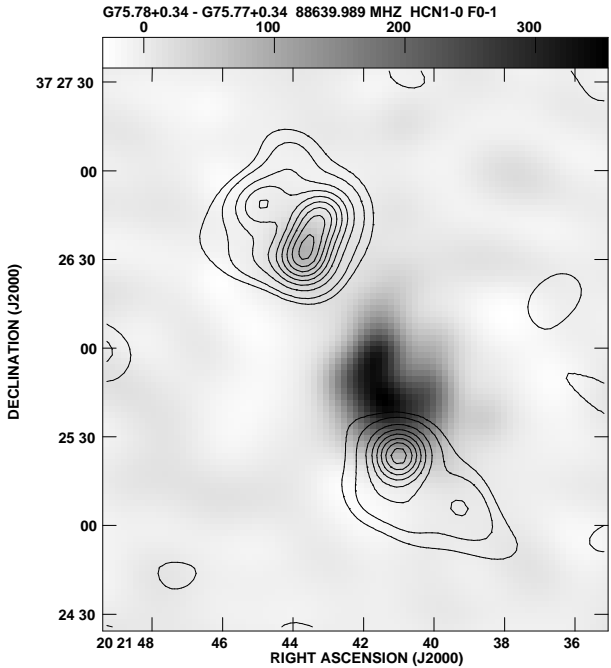}
\caption{HCN J=1-0 F=0-1 flux contours of G75.78+0.34 and G75.77+0.34 
overlaid to the the 3.5\,mm continuum gray scale image}
\label{hcn0-1}
\end{figure}

\begin{figure}
\centering
\includegraphics[scale=0.41]{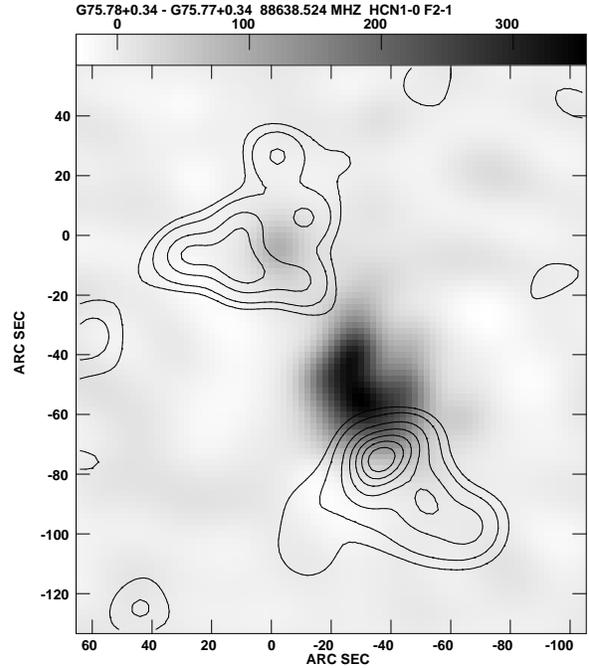}
\caption{Same as Fig.\,\ref{hcn0-1} for the transition F=2-1. The position (0,0) corresponds 
to the location of the ultracompact H\,{\sc ii} region G75.78+0.34.}
\label{hcn2-1}
\end{figure}

\begin{figure}
\centering
\includegraphics[scale=0.41]{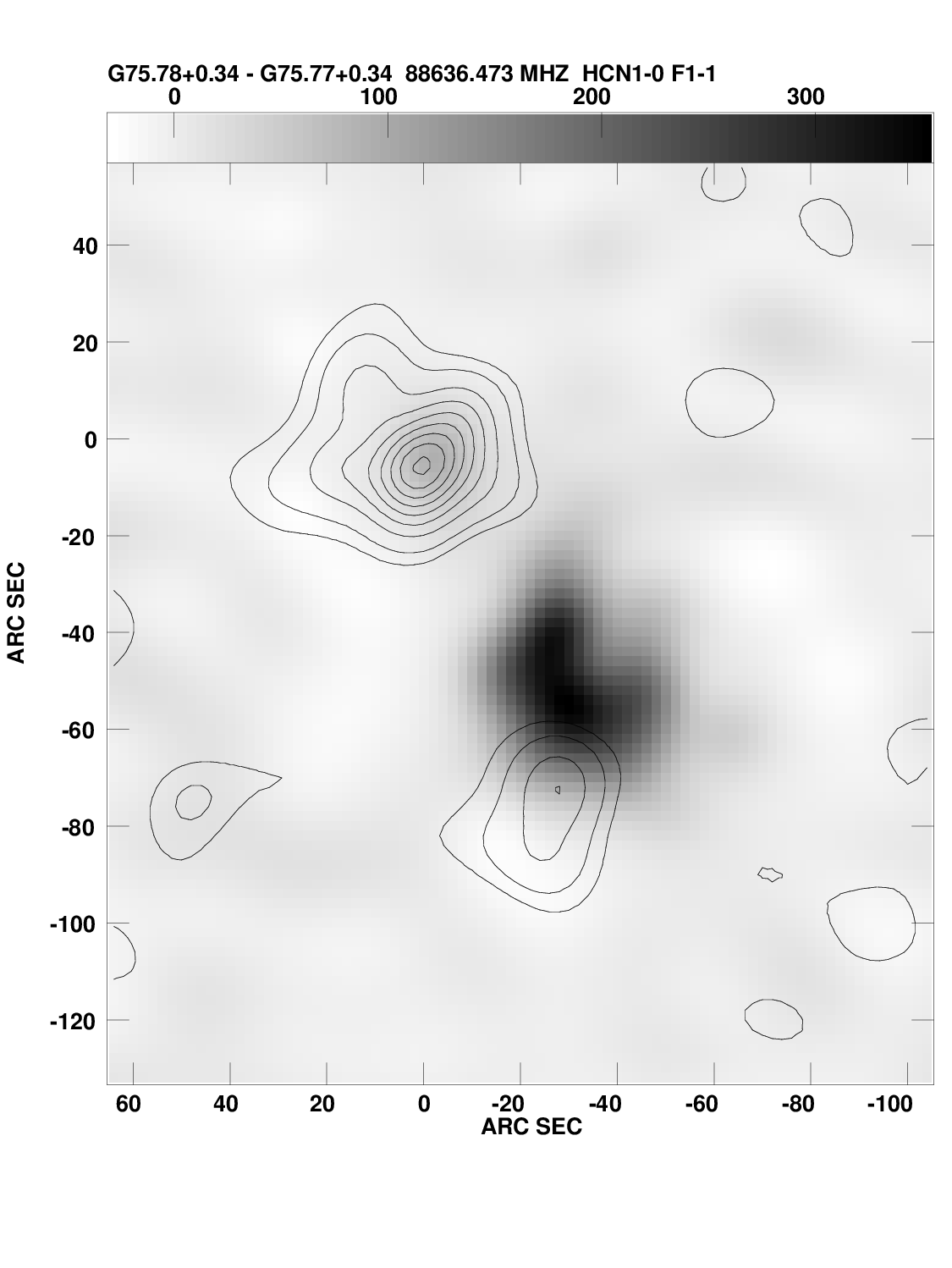}
\caption{Same as Fig.\,\ref{hcn0-1} for the transition F=1-1.}
\label{hcn1-1}
\end{figure}

The HCN line with angular momentum change from J=1-0 presents three hyperfine transitions 
at 88.63042\,GHz (F=1-1), 88.63185\,GHz (F=2-1) and 88.63394\,GHz (F=0-1) \citep[see, e.g.,][]{truong-bach89,afonso98}. 
The high spectral resolution reached by BIMA in D configuration allowed us to separate and construct two-dimensional maps for the 
flux distribution of each hyperfine transition. The resulting flux contours are shown in Figures~\ref{hcn0-1}, \ref{hcn2-1} and \ref{hcn1-1}, for 
 the transitions F=0-1, F=2-1 and F=1-1, respectively. In order to compare the HCN flux distribution with the 
3.5~mm continuum emission we present in these figures the continuum image as a gray scale image.  
Both axis are shown in arcsec units in order to more easily measure the sizes of the extended emission and the position (0,0) 
corresponds to the location of the ultracompact H\,{\sc ii} region G75.78+0.34.  As observed in these figures the HCN emission of G75.78+0.34 for 
the three hyperfine transitions is spatially coincident with the continuum emission, while for G75.77+0.34 it peaks 
at $\alpha=$20h\,21m\,41s and $\delta=$37$^\circ$\,25$^\prime$\,24$^{\prime\prime}$, approximately 25$^{\prime\prime}$ south-west 
from the position of the H\,{\sc ii} region. In contrast to the continuum emission, the HCN emission is 
more extended for G75.74+0.34 than for G75.77+0.34.

A detailed inspection of Figs.~\ref{hcn0-1}, \ref{hcn2-1} and \ref{hcn1-1} reveals distinct morphologies for each 
transition at lower intensity levels. Nevertheless, the global flux distributions are similar to each other and 
 its morphology for G75.78+0.34 can be described as asymmetric presenting  two ``jet-like" structures:
one more extended to the east with size of $\sim50^{\prime\prime}$ (1.3\,pc) and other extending   
up to  $\sim40^{\prime\prime}$  (1.1\,pc) north from the position of the peak intensity.
 A comparison of the HCN emission with the continuum image shows that it is only twice as 
extended as the continuum. G75.77+0.34 also presents
an asymmetric morphology with a sub-structure (with size of $\sim50^{\prime\prime}$) extended to south-east from the peak flux position
 and another with approximately the same size extended to south-west of it.
 In Table~\ref{hcn_flux} we present the measured fluxes and line ratios for each H\,{\sc ii} region.

\begin{figure*}
\centering
\includegraphics[scale=0.81]{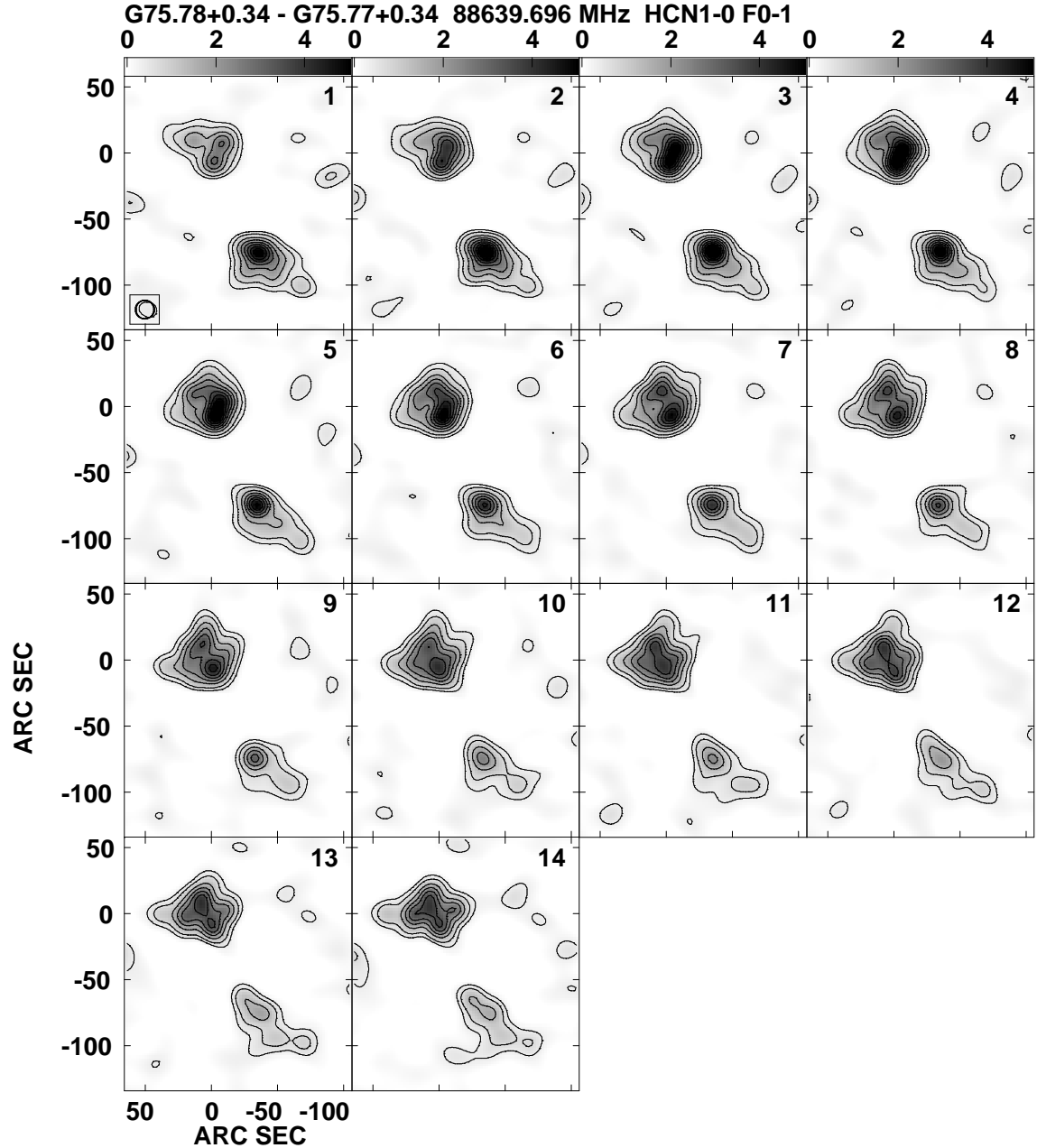}
\caption{Velocity channel panels along the HCN F=0-1  emission-line profile, 
with velocity bin of 0.34\,km\,s${-1}$. The greyscale images show the flux distribution 
at each channel with intensity contours overlaid on a greyscale of the same emission. 
The panel 8 corresponds to the velocity of the peak of the profile and panels from 1 to 7 
represents blueshifts, while panels from 9 to 14 are for redshifts relative to the peak 
of the profile}
\label{channel0-1}
\end{figure*}
\begin{figure*}
\centering
\includegraphics[scale=0.81]{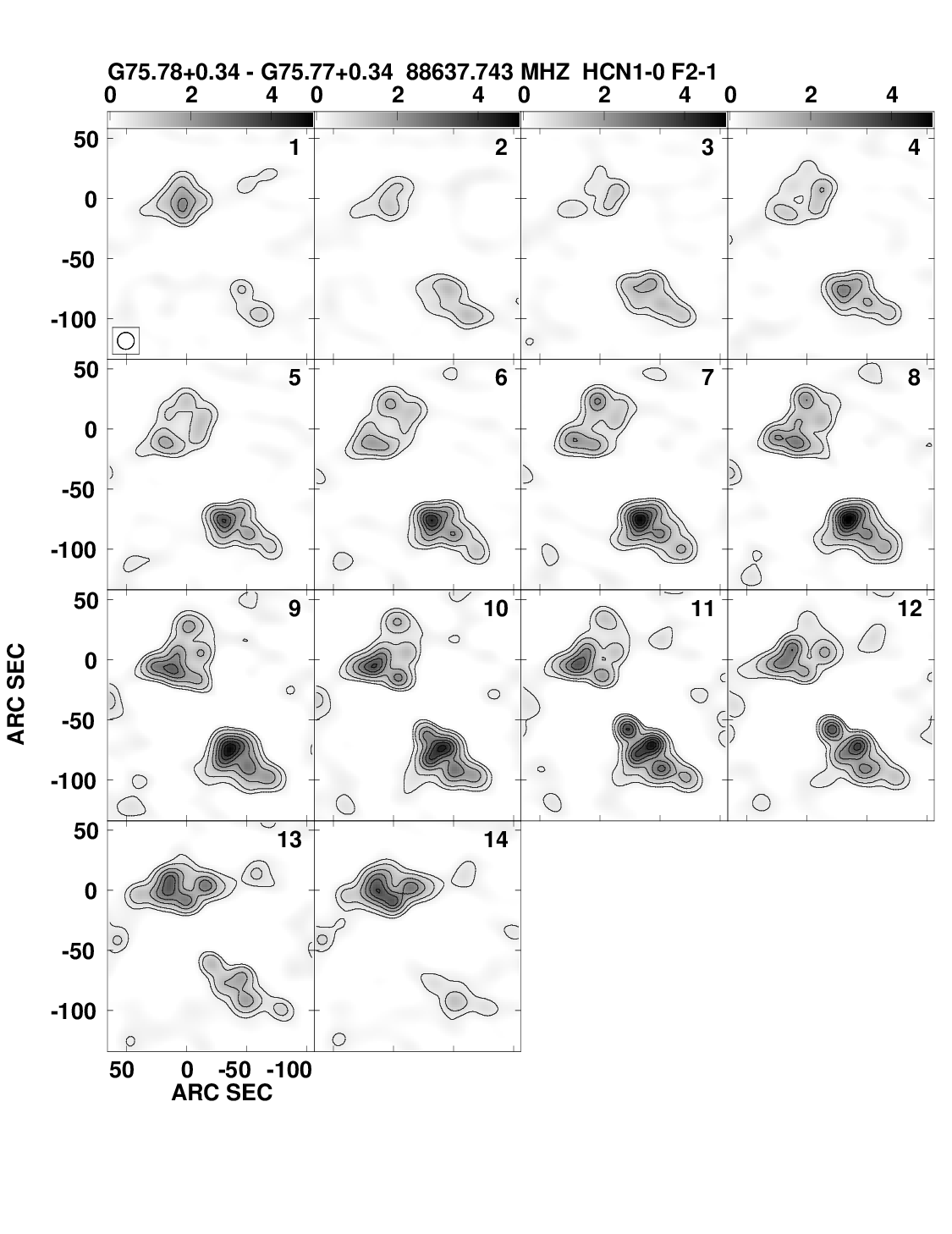}
\caption{Same as Fig.\,\ref{channel0-1} for the transition F=2-1}
\label{channel2-1}
\end{figure*}
\begin{figure*}
\centering
\includegraphics[scale=0.81]{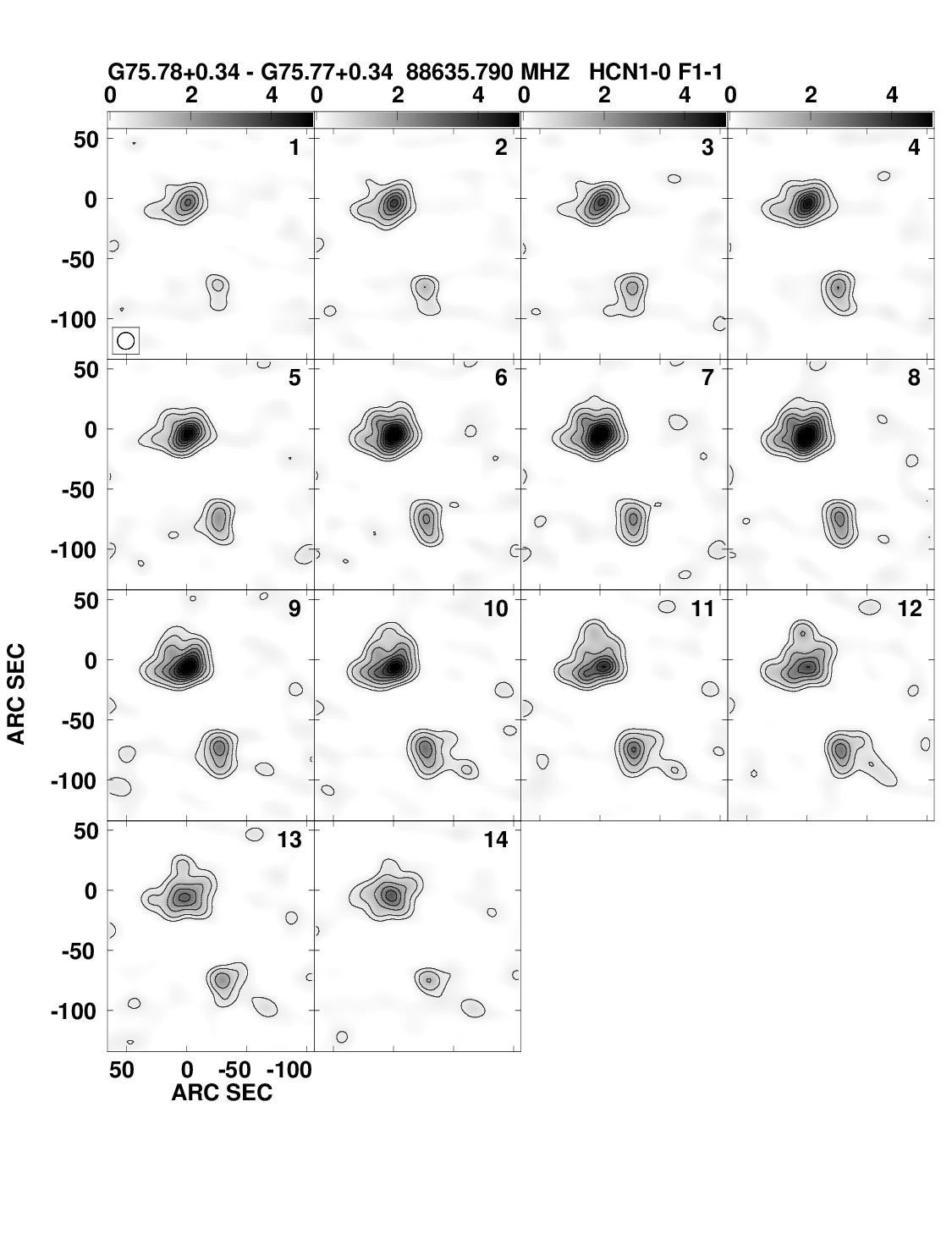}
\caption{Same as Fig.\,\ref{channel0-1} for the transition F=0-1}
\label{channel1-1}
\end{figure*}

In order to investigate the kinematics of the HCN emitting gas
close to both H\,{\sc ii} regions we have constructed velocity channel
maps along each line profile with a velocity bin of 0.34\,km\,s$^{-1}$,
corresponding to our spectral sampling. The resulting velocity
channel maps are shown in Figures \ref{channel0-1}, \ref{channel2-1} and \ref{channel1-1}
 for F=0--1, F=2--1 and F=1--1, respectively. The panel labeled as 
8 is centred at the centroid velocity of each emission-line profile.
 Panels from 1 to 7 correspond to emission from blueshifted gas relative to
the centroid velocity, while panels from 9 to 14 corresponds to emission
from redshifted gas relative to it.

    The velocity channels show
that the G75.78+0.34 and G75.77+0.34  H\,{\sc ii} 
regions present a complex flux distribution and kinematics, with
several knots observed at distinct velocity channels and at distinct locations.
 At the highest blueshifts, usually G75.78+0.34 is dominated by a jet-like
structure to east of the position of the peak flux, also observed in
the HCN fluxes maps; as the velocities approximate to the centroid 
velocity the dominant structure is one elongated to north. At the
 highest redshifts the emission is dominated by gas located in 
a structure oriented east-west (clearly observed in Fig.\ref{channel2-1} panels 13 and 14).
 The G75.77+0.34 velocity channels are dominated a jet-like structure
 extended towards the south-west of the position of peak intensity in all velocities, 
although another emission structure extended to south-east
is also present in the highest blueshifted channel maps
 (more clearly observed at Fig. \ref{channel0-1}).

\section{Discussion} \label{disc}

\subsection{The 3.5\,mm continuum emission}

\citet{shepherd97} used the BIMA array at
configurations B and C to obtain images for the 3.5~mm continuum and
molecular-line emission from H\,{\sc ii} regions of the ON2 molecular
 cloud at spatial resolution of 5$^{\prime\prime}$. It is difficult to compare 
their continuum image with ours due to the poorer spatial resolution used 
in the present work. Nevertheless, we 
can compare the total continuum flux  measured from both images. \citet{shepherd97} 
measured a total flux of 75.4 mJy for G75.78+0.34, which is  about 1.5 times smaller than our
measurement (119 mJy). This difference may be mostly due to the larger aperture used to integrate the flux 
in the present work than those used by \citet{shepherd97}.

    At millimetric wavelengths the continuum emission from H\,{\sc ii}
regions can originate from free-free emission from the 
ionized gas surrounding OB stars and/or by hot dust heated by nearby 
stars, while at higher wavelengths such as at 6~cm the radio continuum 
from such regions is dominated by free-free emission.
 \citet{wood89} measured a flux of 40.4$\pm$0.48\,mJy for 
G75.78+0.34 at 6~cm from VLA observations, which can be used to estimate the contribution of free-free emission to the 3.5\,mm continuum 
under an assumption about the shape of the spectral energy distribution of G75.78+0.34.
 Following \citet{shepherd97},
if we assume a nearly flat spectrum typical of optically thin free-free emission, we would predict a flux of $\sim$40\,mJy 
for the free-free emission at 3.5\,mm and thus the contribution of hot dust would be $\sim$80\,mJy, about two times larger
 than the contribution of the free-free emission. 
In the case of the excess at 3.5\,mm being only due to dust emission, then the peak flux should increase for lower wavelengths, 
which is not observed for G75.78+0.34 as discussed by \citet{shepherd97} based on measurements of the peak fluxes at 3.5\,mm 
and  2.7\,mm (111\,GHz). Nevertheless, the discussion above is based on the assumption that G75.78+0.34 presents a flat spectrum, which 
may not be a good approximation of the spectral energy distribution of the region as pointed out by 
\citet{franco00a}. They  show 
that G75.78+0.34 presents a spectral index $\alpha=+1.4\pm0.1$, suggesting that this ultracompact H\,{\sc ii} is not optically thin, indicating 
a larger importance of the free-free emission to the millimetric continuum.

\citet{garay93} present VLA multi-frequency radio continuum images for several compact H\,{\sc ii} regions, including G75.77+0.34. 
They obtained a total flux of 4.73$\pm$0.03\,Jy at 20\,cm, which is higher than 
those of the  3.5\,mm emission (1.73\,Jy), indicating that  the millimetric continuum emission from this object 
is due to free-free emission in the ionized gas surrounding the O star. This conclusion is true, 
even we assume a negative spectral index of $\alpha=-0.1$, typical for the high frequencies \citep{kurtz05} at optically 
thin regions. 

A comparison between our 3.5\,mm image and the 6\,cm continuum image presented by \citet{matthews73}, for both H\,{\sc ii}
 regions, shows that the two images present similar morphologies, suggesting a same origin for the continuum emission.
 Thus, from this comparison and from the discussion above, 
we conclude that the 3.5\,mm continuum emission from both, G75.78+0.34 and G75.77+0.34, ultracompact 
H\,{\sc ii} is dominated by free-free emission from the ionized gas surrounding the exciting stars. Nevertheless, some contribution 
of dust emission cannot be ruled out for G75.78+0.34.

\subsection{The origin of the HCN emission}

\citet{shepherd96} presented $^{12}$CO(J=1--0) and
 $^{13}$CO(J=1--0) images for ten massive star formation regions
obtained with the National Radio Astronomy Observatory
(NRAO) Kitt Peak 12 m telescope. Their sample included both
G75.77+0.34 and G75.78+0.34 objects (named by the authors as
G75.78 SW and G75.78 NE, respectively). 
A comparison of their CO images with the HCN flux distributions (Figs. \ref{hcn0-1}, \ref{hcn2-1} and \ref{hcn1-1})
 shows that they present similar global morphology.
 In the case of G75.78+0.34 we can also compare the HCN images 
with the CO images presented by \citet{shepherd97}, using higher spatial resolution BIMA observations than 
those reached by \citet{shepherd96}. 
Their image present two ``jet-like" structures, one to east of the 
ultracompact H\,{\sc ii} region and the other to north of it, 
similarly to those observed in the HCN emission,
 although the HCN jets are less extended than
the CO ones and present some knots of higher intensity levels
(more clearly seen in the F=2--1 image), which are not present in
the CO image. The similarity between these images suggests
that the CO and HCN traces the same physical properties. Nevertheless, the
 smaller extension of the jets and the high emission knots in the HCN images indicate
 that its emission traces higher density structures than the ones traced by the 
CO emission, in good agreement with previous studies, which 
found that the J=1--0 transition from the HCN
molecule is a tracer of dense cores with densities in the range $n\sim10^4 - 10^5\,{\rm~cm^{−3}}$ 
 \citep[e.g.][]{cao93,afonso98}.

Besides the HCN flux distribution, its emission origin is one
of the most important questions in the study of molecular cores.
 The populations of the various molecular levels 
are determined by the physical parameters of the gas (temperature, density, velocity) and
the intensity of each hyperfine component may be also affected by radiative
transfer effects. Thus, the HCN hyperfine line ratios  can be used to investigate the physical properties of the 
emitting gas. 
In Local Thermodynamic Equilibrium (LTE) the predicted theoretical values for
the hyperfine line ratios for HCN(J=1--0) are $R_{02}=\frac{F=0-1}{F=2-1}=0.2$ and $R_{12}=\frac{F=1-1}{F=2-1}=0.6$  
 \citep{cirnecharo84,harju89,afonso98}. 
As observed in Table 1 the intensity ratios,
$R_{02}$ and $R_{12}$, differ from the predicted LTE values for both H\,{\sc ii} regions. For
 G75.77+0.34, $R_{02}$ is more than 3 times larger then the predicted ratio, while 
$R_{12}$ is 2.5 smaller. For G75.78+0.34 both ratios are larger than
 the theoretical values -- $R_{02}$ is more than two times larger than
the predicted value, while $R_{12}$ is even larger, reaching a value
 almost six times larger than the theoretical.

 Differences between the intensity ratios observed and predicted
 are commonly reported in the literature and known as the
hyperfine anomaly \citep{harju89,gonzalez-alfonso93,cao93,afonso98,kim02}. 
Two scenarios have been proposed to explain this anomaly. The
  thermal model developed by \citet{guilloteau81} suggests that the overlap of the J=2--1 
hyperfine transitions overpopulates the state J=1, F=2, and thus the line J=1--0, F=2--1
  grows relative to the other lines. With the growing  temperature, the ratios $R_{02}$ and $R_{12}$
 become smaller than the LTE values. This model provides a reasonable explanation for the observed 
ratios in hot clouds. The second scenario, proposed by \citet{cirnecharo84}, 
 suggests that the relative intensities of HCN hyperfine
 transitions are formed by scattering of the radiation emitted
 from the cloud core to the surrounding envelope. The optically-thick lines (F=2--1 and F=1--1) are scattered more often than the
 optically-thin line (F=0--1), and thus the line F=0--1 is enhanced
 relative to the other lines.

    The scattering scenario have been invoked to explain the intensity 
ratios observed in ultracompact H\,{\sc ii} regions \citep[e.g.][]{kim02,harju89} and could explain 
the emission from G75.78+0.34 and G75.77+0.34. This suggestion is supported by the distinct flux distributions 
observed for each hyperfine transition from both H\,{\sc ii} regions (see Figs.\,\ref{hcn1-1}, \ref{hcn2-1} and \ref{hcn0-1}). 
For G75.78+0.34, the flux distribution for F=0--1 is the most concentrated, followed by F=1--1 and F=2--1, 
for which the scattering of the radiation emitted from the cloud core is more important.
 In the case of G75.77+0.34 the $R_{12}$ is smaller than the 
predicted for LTE, which favors the thermal model in which the intensity of F=2--1 grows relative to the other lines.
 Thus, numerical models and  higher spatial resolution 
line ratio maps are necessary to properly distinguish between both scenarios for both H\,{\sc ii} regions. 

\subsection{Molecular outflows}

As discussed in Sec.\,\ref{hcn_flux}, the HCN(J=1--0) and $^{12}$CO(J=1--0)
flux distributions are similar for G75.78+0.34. From the velocity
channels (Figs.\,\ref{channel0-1}, \ref{channel2-1}, \ref{channel1-1}) we 
can investigate the kinematics of the HCN emitting gas and compare it with the CO 
kinematics presented by \citet{shepherd96} and \citet{shepherd97}. 
These works found bipolar molecular outflows associated with the ultracompact H\,{\sc ii} region, with
the blueshifted emitting gas presenting sub-structures elongated
to north and to east from the peak flux position and the 
redshifted gas being more elongated to south-west of it. A detailed
analysis of the HCN velocity channels reveals that the blueshifted 
HCN emission is dominated by the two ``jet-like" structures described above.
The peak of 
redshifted HCN emission is a bit displaced to south relative to the
peak position of the blueshifted emission and the highest velocity
channels present an additional structure extended to south-west.
These kinematic components are similar to those observed in 
$^{12}$CO(J=1--0)  indicating that at least part of the 
HCN emitting gas follow the bipolar molecular outflows 
observed in CO.

     In the case of G75.77+0.34 there are no velocity channel
maps with similar resolution to ours for the CO emission in
the literature and thus a comparison between the HCN and CO
kinematics is not possible here. The flux distribution of all channel maps are 
similar, with exception of the highest velocity channels, which show 
three knots oriented to south-west of the position of the peak emission, indicating 
that dense molecular outflows are less important for this object than for the case of 
G75.78+0.34.

\section{Conclusions} \label{conc}

We analyzed 3.5~mm continuum and HCN(J=1--0) line emission
from the H\,{\sc ii} regions G75.78+0.34 and G75.77+0.34 from interferometric
 observations obtained with the BIMA array in D configuration. We
present for the first time images for the HCN(J=1--0) for both
H\,{\sc ii} regions. The main results of this work are:

\begin{itemize}

\item The the 3.5~mm continuum emission is consistent with the free-free emission from the ionized gas from the H\,{\sc ii} regions.
    However, the contribution of emission from hot dust cannot
    be totally discarded for G75.78+0.34.

\item The flux distributions for the HCN(J=1--0) hyperfine lines
    present similar structures than those observed in $^{12}$CO(J=1--0) images, suggesting 
    both gases traces the same physical conditions. Some knots of high intensity are present only in the 
    HCN images suggesting the presence of high density regions not observed in CO. 

\item The analysis of the HCN(J=1--0) hyperfine intensity ratios
    reveals that they are different than those predicted theoretically for LTE, 
    probably due to scattering of radiation processes.

\item The kinematical analysis reveal that the HCN emitting gas
    follows the bipolar molecular outflows observed in CO for G75.78+0.34, while  
   for G75.77+0.34 the outflows seems to be less important.
\end{itemize}

\section*{Acknowledgments}
We thank the referee for valuable suggestions which helped to improve the present paper. 
   The BIMA radio observatory is a consortium among University of Maryland, University
of Illinois and UCLA on behalf of NSF, USA. 
This work has been partially supported by the Brazilian institution CAPES.

\label{lastpage}

\end{document}